%% file: main.tex
\documentclass[sigconf,review=false]{acmart}

\AtBeginDocument{%
  \providecommand\BibTeX{{%
    \normalfont B\kern-0.5em{\scshape i\kern-0.25em b}\kern-0.8em\TeX}}}

\renewcommand\footnotetextcopyrightpermission[1]{} 
\setcopyright{none}

\usepackage{amsmath,amssymb,amsfonts}
\usepackage{algorithm}
\usepackage[noend]{algpseudocode}
\usepackage{textcomp}
\usepackage{xcolor}
\usepackage{soul}
\usepackage{booktabs}
\usepackage[utf8]{inputenc}
\usepackage[T1]{fontenc}
\usepackage{microtype}
\usepackage{rotating}
\usepackage{graphicx}
\usepackage{paralist}
\usepackage{tabularx}
\usepackage{multicol}
\usepackage{multirow}
\usepackage{pbox}
\usepackage{enumitem}	
\usepackage{colortbl}
\usepackage{pifont}
\usepackage{xspace}
\usepackage{url}
\usepackage{tikz}
\usepackage{tabularx}
\usepackage{fontawesome}
\usepackage{lscape}
\usepackage{listings}
\usepackage{color}
\usepackage{anyfontsize}
\usepackage{comment}
\usepackage{soul}
\usepackage{multibib}
\include{macro}

\newcommand{\HIDE}[1]{}

\begin{document}
\pagestyle{plain}
\title{Taxonomy of Real Faults in Deep Learning Systems}

\author{Nargiz Humbatova}
\email{nargiz.humbatova@usi.ch}
\affiliation{
  \institution{Universit\`a della Svizzera Italiana (USI)}
  \city{Lugano}
  \state{Switzerland}
}

\author{Gunel Jahangirova}
\email{gunel.jahangirova@usi.ch}
\affiliation{
  \institution{Universit\`a della Svizzera Italiana (USI)}
  \city{Lugano}
  \state{Switzerland}
}

\author{Gabriele Bavota}
\email{gabriele.bavota@usi.ch}
\affiliation{
  \institution{Universit\`a della Svizzera Italiana (USI)}
  \city{Lugano}
  \state{Switzerland}
}

\author{Vincenzo Riccio}
\email{vincenzo.riccio@usi.ch}
\affiliation{
  \institution{Universit\`a della Svizzera Italiana (USI)}
  \city{Lugano}
  \state{Switzerland}
}

\author{Andrea Stocco}
\email{andrea.stocco@usi.ch}
\affiliation{
  \institution{Universit\`a della Svizzera Italiana (USI)}
  \city{Lugano}
  \state{Switzerland}
}

\author{Paolo Tonella}
\email{paolo.tonella@usi.ch}
\affiliation{
  \institution{Universit\`a della Svizzera Italiana (USI)}
  \city{Lugano}
  \state{Switzerland}
}

\renewcommand{\shortauthors}{Jahangirova and Humbatova, et al.}

\begin{abstract}
The growing application of deep neural networks in safety-critical domains makes the analysis of faults that occur in such systems of enormous importance. In this paper we introduce a large taxonomy of faults in deep learning (DL) systems. We have manually analysed 1059 artefacts gathered from GitHub commits and issues of projects that use the most popular DL frameworks (TensorFlow, Keras and PyTorch) and from related Stack Overflow posts. Structured interviews with 20 researchers and practitioners describing the problems they have encountered in their experience have enriched our taxonomy with a variety of additional faults that did not emerge from the other two sources. Our final taxonomy was validated with a survey involving an additional set of 21 developers, confirming that almost all fault categories (13/15) were experienced by at least 50\% of the survey participants.
\end{abstract}

%


\maketitle

\input{introduction}
\input{related}

\input{methodology}
\input{results}
\input{discussion}
\input{threats}

\input{conclusion}

\bibliographystyle{ACM-Reference-Format}
\bibliography{main}

\end{document}

%% file: introduction.tex

\section{Introduction} \label{sec:intro}

Deep Learning (DL) is finding its way into a growing number of areas in science and industry. Its application ranges from supporting daily activities such as converting voice to text, translating texts from one language to another, to much more critical tasks such as fraud detection in credit card companies, diagnosis and treatment of diseases in medical field, autonomous driving of vehicles. The increasing dependence of safety-critical systems on DL networks makes the types of faults that can occur in such systems a crucial topic. However, the notion of fault in DL systems is  more complex than in traditional software. In fact, the code that builds the DL network might be bug free, but the network might still deviate from the expected behaviour due to faults introduced in the training phase, such as the misconfiguration of some learning parameters or the use of an unbalanced/non-representative training set.

The goal of this paper is to build a \textit{taxonomy} of real faults in DL systems. Such a taxonomy can be useful to aid developers  avoiding common pitfalls or can serve as a checklist for testers, motivating them to define test scenarios that address specific fault types. The taxonomy could also be used for fault seeding, as resemblance with real faults is an important feature for artificially injected faults.

A taxonomy is mainly a classification mechanism \cite{Usman:2017}.  According to Rowley and Farrow \cite{Rowley:2017}, there are two main approaches to classification: enumerative and faceted. In \textit{enumerative classification}, the classes are predefined. However, it is difficult to enumerate all classes in immature or evolving domains, which is the case of DL systems. Therefore, using a vetted taxonomy of faults \cite{Beizer:1984} would not be appropriate for our purpose. In contrast, in \textit{faceted classification} the emerging traits of classes can be extended and combined. For this reason we used faceted classification, \ie we created the categories/subcategories of our taxonomy in a bottom up way, by analysing various sources of information about DL faults.

Our methodology is based on the manual analysis of unstructured sources and interviews. We started by manually analysing 477 Stack Overflow (SO) discussions, 271 issues and pull requests (PRs), and 311 commits from GitHub repositories, in which developers discuss/fix issues encountered while using three popular DL frameworks. The goal of the manual analysis was to identify the root cause behind the problem. The output of this step was the first hierarchical taxonomy of faults related to the usage of DL frameworks. Then, two of the authors interviewed 20 researchers and practitioners to collect their experience on the usage of DL frameworks. All the interviews were taped and transcribed, allowing an open coding procedure among all authors, by which we identified the categories of faults mentioned by the interviewees. This allowed to complement our preliminary taxonomy and to produce its final version.

To validate our final taxonomy we have conducted a survey to which an additional set of 21 researchers/practitioners have responded. In the survey, we included the categories of the taxonomy along with a description of the types of faults they represent, and asked the participants to indicate whether these faults have been encountered in their prior experience when developing DL systems. Most faults (13/15) were experienced by 50\% or more of the participants and no fault category remained non-validated (the least frequent category was confirmed by 24\% participants).

The main contribution of this paper is the validated taxonomy of real faults in DL systems. More specifically, to our knowledge this is the first work that includes interviews with developers on the challenges/problems related to the development of DL systems. Without such interviews, 2 inner nodes and 27 leaf nodes of the taxonomy would be completely missed. Several other nodes are highly represented in interviews, but they appear quite rarely in the other analysed artefacts. 

\textbf{Structure of the paper}. \secref{sec:related} gives an overview of related work. \secref{sec:methodology} describes the methodology used to collect faults, to build the taxonomy and to validate it. The final version of the taxonomy, along with the description of its categories and the results of the validation survey are presented in \secref{sec:results}. \secref{sec:discussion} contains a discussion of our findings, while \secref{sec:threats} reviews our threats to validity. Finally, \secref{sec:conclusion} concludes the paper. 

%% file: related.tex

\section{Related Work} \label{sec:related}

One of the first papers considering faults specifically in Machine Learning (ML) systems is the empirical study by Thung \etal \cite{Thung:2012}. The authors manually labeled 500 bug reports and fixes from the bug repositories of three open source projects (Apache Mahout, Lucene, and OpenNLP) to enable their classification into the categories proposed by Seaman \etal \cite{Seaman:2008}. Descriptive statistics were used to address research questions such as how often the bugs appear, how severe the bugs are and how much effort is put into their resolution. 

A similar study was published in 2017 by Sun \etal \cite{Sun:2017}. The authors examined bug patterns and their evolution over time. As a subject of the study the authors considered three ML projects from GitHub repositories (Scikit-learn, Caffe and Paddle). The collected issues have been organised into 7 categories. Manual analysis of 329 successfully closed bug reports allowed the authors to assess the fault category, fix pattern, and effort invested while dealing with a bug. 

The main difference between these two works and our study is that they analysed  bugs \textit{in} the frameworks themselves while we focus on the faults experienced when building DL systems that \textit{use} a specific framework.

Zhang \etal \cite{Zhang:2018} studied a number of DL applications developed using TensorFlow. They collected information about 175 TensorFlow related bugs from SO and Github. Manual examination of these bugs allowed the authors to determine the challenges developers face and the strategies they use to detect and localise faults. The authors also provide some insight into the root causes of bugs and into the consequences that bugs have on the application behaviour. The authors were able to classify their dataset into seven general kinds of root causes and four types of symptoms. 

In our study, we analyse DL applications that use the most popular DL frameworks \cite{framework_data}, TensorFlow, PyTorch and Keras, not just the former. The popularity of these three frameworks (in particular, Keras) and their strong prevalence over other similar products allows us to consider them as representative of the current situation in the field. Another methodological difference lies in the mining of the SO database. Zhang \etal considered SO questions under the constraint that they had at least one answer, while we analysed only questions with an \textit{accepted} answer, to be sure that the fault was investigated in depth and solved. As for GitHub, Zhang \etal used only 11 projects to collect the faults. After a complex filtering and cleaning process, we were able to use 564 projects. For further comparison, Zhang \etal found in total 175 bugs, which include those that we discarded as generic (\ie non DL specific), while our taxonomy bears 375 DL specific faults in total. It is difficult to compare the overall number of analysed artefacts as such statistics are not reported in Zhang \etal's paper. Last but not least, we decided not to limit our analysis to just SO and Github: we included interviews with researchers and practitioners, which revealed to be a key contribution to the taxonomy. A detailed comparison between our and Zhang \etal's taxonomy is reported in \secref{sec:discussion}.

Another work worth mentioning is a DL bug characterisation by Islam \etal \cite{Islam:2019}. The aim of the authors is to find what types of bugs are observed more often and what are their causes and impacts. They also investigated whether the collected issues follow a common pattern and how  this pattern evolved over time. The authors studied a number of SO and GitHub bugs related to five DL frameworks: Theano, Caffe, Keras, TensorFlow and PyTorch. To perform their analysis, the authors labeled the dataset according to a classification system adapted from the 1984 work by Beizer \cite{Beizer:1984}. For the categorisation of the bug causes, the authors adopted the list of root causes from Zhang \etal \cite{Zhang:2018}. Differently from us, Islam \etal did not have the aim of building a comprehensive fault taxonomy. Instead, they performed an analysis of various fault patterns and studied the correlation/distribution of bugs in different frameworks, reusing existing taxonomies available in the literature.

%% file: methodology.tex

\section{Methodology} \label{sec:methodology}

\subsection{Manual Analysis of Software Artefacts} 
\label{sub:bug_analysis}
To derive our initial taxonomy we considered the three most popular DL frameworks~\cite{framework_data}, TensorFlow, Keras and PyTorch. We manually analysed four sources of information: commits, issues, pull requests (PRs) from GitHub repositories using TensorFlow, Keras, or PyTorch, and SO discussions related to the three frameworks. 

\subsubsection{Mining GitHub}
\label{sub:miningGH}
We used the GitHub search API \cite{gh-search} to identify repositories using the three DL frameworks subject of our study. The API takes as an input a search string and fetches source code files from GitHub repositories that match the search query. For example, in Python, TensorFlow can be imported by using the statement \texttt{import tensorflow as tf}. Thus, we used the search string ``tensorflow'' to identify all Python files using TensorFlow. Clearly, this may result in a number of false positives, since the string ``tensorflow''  may be present  inside a source file for other reasons (\eg as part of a String literal printed on screen). However, the goal of this search is only to identify \textit{candidate} projects using the three frameworks, and false positives are excluded in subsequent steps.The search strings we defined are ``tensorflow'', ``keras'', and ``torch''. We limited the search to Python source files using the \texttt{language:python} argument.

While using the GitHub search API, a single request can return 1,000 results at most. To overcome this limitation, we generated several requests, each having a specific size range. We used the \texttt{size:min..max} argument to retrieve only files within a specific size range. In this way, we increased the number of returned results to up 1,000 $\times$ $n$, where $n$ is the number of considered size ranges. For each search string, we searched for files having a size ranging from 0 to 500,000 bytes, with a step of 250 bytes. Overall, we generated 6,000 search requests, 2,000 for each framework. 

For each retrieved Python file we identified the corresponding GitHub repository, and we extracted relevant attributes such as: number of commits, number of contributors, number of issues/PRs, number of stars \cite{stars}, and number of forks \cite{forks}.
Then, we excluded: (i) personal repositories, classified as those having less than five contributors; (ii) inactive repositories, \ie having no open issues; (iii) repositories with trivial history, \ie having less than 100 commits; and (iv) unpopular repositories, that we identified as those with less than 10 stars and 10 forks.

\HIDE{
\begin{itemize}[leftmargin=*]
\item {\bf Personal.} Repositories having less than five (5) contributors, as an attempt to remove from our dataset personal, toy projects.

\item {\bf Inactive.} Repositories having no open issues, as a symptom of their scarce activity.

\item {\bf Trivial history.} Repositories having less than 100 commits, to ensure that the projects had enough commits for our analysis.

\item {\bf Unpopular.} Repositories with less than 10 stars and 10 forks, to avoid likely irrelevant projects.
\end{itemize}
}

Such a process resulted in the selection of 151 TensorFlow projects, 237 Keras projects, and 326 PyTorch projects. Then, one of the authors checked the selected repositories with the goal of excluding tutorials, books, or collections of code examples, not representing real software systems used by developers in practice, and false positives (\ie projects containing the search strings in one of their Python files but not actually using the related framework). This process left us with 121 TensorFlow, 175 Keras, and 268 PyTorch projects.

For each of the retained 564 projects, we collected issues, PRs, and commits likely related to fixing problems/discussing issues. For issues and PRs, we used the GitHub API to retrieve all those labelled as either \texttt{bug}, \texttt{defect}, or \texttt{error}. For  commits, we mined the change log of the repositories to identify all those having a message that contained the patterns \cite{Fischer:icsm2013}:  (``fix'' or ``solve'') and (``bug'' or ``issue'' or ``problem'' or ``defect'' or ``error''). 

Then, for each framework, we selected a sample of 100 projects for manual analysis. Instead of applying a random selection, we selected the ones having the highest number of issues and PRs. For frameworks for which less than 100 projects had at least one relevant issue/PR, we selected the remaining projects sorting them by the number of relevant commits (\ie commits matching the pattern described above). The 100 selected projects account for a total of 8,577 issues and PRsand 28,423 commits. 

Before including these artefacts in our study, we manually inspected a random sample of 100 elements and found many (97) false positives, \ie issues/PRs/commits that, while dealing with fault-fixing activities, were unrelated to issues relevant to the usage of the underlying DL framework (\ie were about generic programming bugs). Thus, we decided to perform a further cleaning step to increase the chance of including relevant documents in the manual analysis. We defined a vocabulary of relevant words related to DL (\eg ``epoch'', ``layer''), and excluded all artefacts that did not contain any of these words. Specifically, we extracted the complete list of 11,986 stemmed words (\ie ``train'', ``trained'', and ``training'' were counted only once as ``train'') composing the vocabulary of the mined issues, PRs and commits. For commits, we searched for the relevant words in their commit note, whereas for issues and PRs we considered title, description, and all comments posted in the discussion. We sorted the resulting words by frequency (\ie number of artefacts in which they appear), and we removed the long tail of words appearing in less than 10 artefacts. This was done to reduce the manual effort needed to select  the words relevant for DL from the resulting list of words. Indeed, even assuming that one of the automatically discarded rare words were relevant for DL, this would have resulted in missing at most nine documents in our dataset. The remaining 3,076 words have been manually analysed: we split this list into five batches of equal size, and each batch was assigned to one author for inspection, with the goal of flagging the DL relevant words. All flagged words were then discussed in a meeting among all authors in which the final list of 105 relevant words was defined. The list is available in our replication package \cite{replication}, and includes words such as \emph{layer}, \emph{train}, \emph{tensor}. After excluding all artefacts  not containing at least one of the 105 relevant words, we obtained the final list of commits (1,981), and of issues/PRs (1,392). 

\subsubsection{Mining Stack Overflow}
\label{sub:miningSO}
We used StackExchange Data Explorer \cite{dataexplorer} to get the list of SO posts related to TensorFlow, Keras and PyTorch. StackExchange Data Explorer is a web interface that allows the execution of SQL queries on data from Q\&A sites, including SO. For each framework, we created a query to get the list of relevant posts. We first checked if the name of a framework is indicated in the post's tags. Then, we filtered out posts which contained the word "how'', "install'' or "build'' in their title, to avoid general how-to questions and requests for installation instructions. We also excluded posts that did not have an accepted answer, to ensure that we consider only questions with a confirmed solution. As a result, we obtained 9,935 posts for Tensorflow, 3,116  for Keras, and 1,007 for PyTorch. We ordered the results of each query by the number of times the post has been viewed. Then, to select the posts that may be addressing the most relevant faults, we selected the top 1,000 most viewed posts (overall, 3,000 posts). 

\subsubsection{Manual Labelling}
\label{sub:manual}
The data collected from GitHub and SO was manually analysed by all authors following an open coding procedure~\cite{Seaman:1999}. The labelling process was supported by a web application that we developed to classify the documents (\ie to describe the reason behind the issue) and to solve conflicts between the authors. Each author independently labelled the documents assigned to her by defining a descriptive label of the fault. 
During the tagging, the web application shows the list of labels already created, which can be used by an evaluator should an existing label apply to the fault under analysis. Although, in principle, this is against the notion of open coding, little is still known on DL faults, and the number of possible labels may grow excessively. Thus, such a choice was meant to help coders use consistent naming without introducing substantial bias. 

The authors followed a rigorous procedure for handling special and corner cases. Specifically, (i) We marked as a \emph{false positive} any analysed artefact that either was not related to any issue-fixing activity or happened to be an issue in the framework itself rather than in a DL system. (ii) If the analysed artefact concerned a fix, but the fault itself was not specific to DL systems, being rather a common programming error (\eg wrong stopping condition in a \texttt{for} loop), we marked it as \emph{generic}. (iii) If the artefact was related to issue-fixing activities and it was specific of DL systems, but the evaluator was not able to trace back the root cause of the issue, the \emph{unclear} label was assigned.

When inspecting the documents, we did not limit our analysis by reading only specific parts of the document. Instead, we looked at the entire SO discussions, as well as the entire discussions and related code changes in issues and PRs. For commits, we looked at the commit note as well as at the code diff.

In cases where there was no agreement between the two evaluators, the document was automatically assigned by the web platform to an additional evaluator. In case of further disagreement between the three evaluators, conflicts were discussed and solved within dedicated meetings among all authors. 

\begin{table}[ht]
\scriptsize
\caption{Manual Labelling Process\vspace{-0.3cm}}
\resizebox{\linewidth}{!}{
\begin{tabular}{cccccc} \hline
\multirow{3}{*}{\bf Round} & \bf Analyzed & \multirow{3}{*}{\bf Conflicts} & \bf Relevant & \bf New Inner & \bf New Leaf\\

& \bf Artifacts & & \bf to DL & \bf Categories &  \bf Categories\\

&  & &  & \bf(1st lvl./2nd lvl./3d lvl.) & \\
\hline
1 & 29   & 3   & 5   & -/-/- & 5\\
2 & 110 & 11 & 16  & -/-/- & 11\\
3 & 134 & 8  & 14  & -/-/- & 5\\
4 & 126 & 11 & 20 & 5/7/7 & 10\\ 
5 & 345 & 31 & 46 & 0/2/1 & 21\\
6 & 315 & 47 & 48 & 0/0/0 & 13\\
\hline
\bf Sub Total & \bf 1059 & \bf 111 & \bf 149 & \bf 5/9/8 & \bf 65\\
\hline
Interviews & 297 & 6 & 226 & 0/2/0 & 27\\
\hline
\bf Total & \bf 1356 & \bf 117 & \bf 375 & \bf 5/11/8 & \bf 92\\
\hline
\end{tabular}
}
\vspace{-0.2cm}
\label{tab:labeling}
\end{table}

The labelling process involved six rounds, each followed by a meeting among all authors to discuss the process and solve conflicts. \tabref{tab:labeling} reports statistics for each of the six rounds of labelling, including: (i) the number of artefacts analysed by at least two authors; (ii) the number of artefacts for which conflicts were solved in the following meeting; (iii) the number of artefacts that received a label identifying faults relevant for DL systems; and (iv) the number of new top/inner/leaf categories in the taxonomy of faults. 

In the first three rounds we defined a total of 21 (5+11+5) leaf categories grouping the 35 DL-relevant artefacts. With the growing number of categories in round four, we started creating a hierarchical taxonomy (see \figref{fig:taxonomy}), with inner nodes grouping similar ``leaf categories''. 

\tabref{tab:labeling} shows the number of inner categories created in the fourth, fifth, and sixth round organised by level (1st level categories are the most general). We kept track of the number of inner categories produced during our labelling process and decided to stop the process when we reached saturation for such inner categories, \ie when a new labelling round did not result in the creation of any new inner categories in the taxonomy.

In the last two rounds, we increased the number of labels assigned to each author. We opted for a longer labelling period because the process was well-tuned and there was no need for regular meetings.  
Overall, we labelled 1,059 documents, and 111 (10.48\%) of them required conflict resolution in the open discussion meetings. 

\subsection{Developer Interviews}
\label{sub:interviews}

Unlike traditional systems, DL systems have unique characteristics, as their decision logic is not solely implemented in the source code, but also determined by the training phase and the structure of the DL model (\eg number of layers). While SO posts and GitHub artefacts are valuable sources of information for our study, the nature of these platforms limits the issues reported to mostly code-level problems, hence possibly excluding issues encountered \eg during model definition or training. 
To get a more complete picture, we have interviewed 20 researchers/practitioners with various backgrounds and levels of expertise, focusing on the types of faults encountered during the development of DL-based systems.

\subsubsection{Participant Recruitment}
\label{sec:int_recr}

To acquire a balanced and wide view on the problems occurring in the development of real DL systems, we involved two groups of developers: researchers and practitioners. In the former group, we considered PhD students, Post-Docs and Professors engaged in frequent usage of DL as a part of their research. The second group of interviewees included developers working in industry or freelancers, for whom the development of DL applications was the main domain of expertise.

We exploited three different sources to attract participants. First, we selected candidates from personal contacts. This resulted in a list of 39 developers, 20 of whom were contacted via e-mail. We received 12 positive responses from 9 researchers and 3 practitioners. To balance the ratio between researchers and practitioners, we referred to other two sources of candidates. One of them was SO, whose top answerers are experienced DL developers with proven capability to help other developers solve recurring DL problems. To access the top answerers we referred to statistics associated with the tag that represents each of the three frameworks we study. We used the `\textit{Last 30 Days}' and `\textit{All Time}' categories of top answerers and extracted the top 10 answerers from both categories for each tag (DL framework), resulting in 60 candidates in total. As there is no built-in contact form on SO, it was not possible to get in touch with all of the 60 shortlisted users. We managed to locate email addresses for 17 of them from links to personal pages that users left on their SO profiles. From 17 of the contacted users, we received 6 responses, of which 4 were positive (divided into 3 practitioners and 1 researcher). 

The other source was Upwork \cite{upwork}, a large freelancing platform. We created a job posting with the description of the interview process on the Upwork website. The post was restricted to invited public. The invited candidates were selected according to the following criteria: (i) a candidate profile should represent an individual and not a company, (ii) the candidate's job title should be DL-related, (iii) the candidate's recently completed projects should mostly be DL-related, (iv) Upwork success rate of the candidate should be higher than 90\% and (v) the candidate should have earned more than 10,000 USD on the Upwork platform. From 23 invitations sent, 5 candidates accepted the offer, but one of them was later excluded, being a manager of a team of developers, and not a developer herself. 

Overall, the participant recruitment procedure left us with 20 successfully conducted interviews, equally divided among researchers and practitioners (10 per group). 

Detailed information on the participants' DL experience is available in our replication package \cite{replication}. For what concerns the `overall coding experience', among the interviewed candidates the lowest value is 2.5 years and the highest is 20 years (median=5.4). As for the DL-specific `relevant experience', the range is from 3 months to 9 years (median=3). 

The interviewees reported to use Python as a main programming language to develop DL applications, with a few mentions to Matlab, R, Java, Scala, C++ and C\#. Concerning the usage of DL frameworks, TensorFlow was mentioned 12 times, Keras 11, and PyTorch 8 times. The domains of expertise of the interviewees cover a wide spectrum, from Finance and Robotics to Forensics and Medical Imaging.

\subsubsection{Interview Process} \label{sec:interview_process}

Since we are creating a taxonomy from scratch, rather than classifying issues and problems into some known structure, the interview questions had to be as generic and open-ended as possible. We opted for a \textit{semi-structured interview} \cite{Seaman:1999}, which combines open-ended questions (to elicit unexpected types of information) with specific questions (to keep the interview within its scope and to aid interviewees with specific questions). 

In semi-structured interviews the interviewer needs to improvise new questions based on the interviewee's answer, which might be a challenging task. Therefore, it might be useful to have an additional interviewer who can ask follow-up questions and support the primary interviewer in case of need. For this reason our interviews were conducted by two authors simultaneously, but with different roles: one led the interview, while the other asked additional questions only when  appropriate. The work by Hove \etal \cite{Hove:2005} shows that half of the participants in their study talked much more when the interviews were conducted by two interviewers instead of one. This was the case also in our experience, as in all the interviews the second interviewer asked at least two additional questions.

After collecting information about the interviewees' general and DL-specific programming experience, we proceeded with the questions from our interview guide \cite{replication}. Our first question was very general and was phrased as ``\emph{What types of problems and bugs have you faced when developing ML/DL systems?}''. Our aim with this question was to initiate the topic as open-ended as possible, allowing the interviewees to talk about their experience without directing them to any specific kind of faults. Then, we proceeded with more specific questions, spanning among very broad DL topics, such as training data, model structure, hyperparameters, loss function and hardware used. We asked the interviewees if they ever experienced issues and problems related to these topics and then, if the answer was positive, we proceeded with more detailed questions to understand the related fault.

All of our interviews were conducted remotely (using Skype), except one which was conducted in person. The length of the interviews varied between 26 and 52 minutes, with an average of 37 minutes. For each interview one of the two interviewers was also the transcriber. For the transcription, we used Descript \cite{descript}, an automated speech recognition tool that converts audio/video files into text. After the automated transcription was produced, the transcriber checked it and did manual corrections in case of need.
 
\subsubsection{Open Coding}

To proceed with open coding of the transcribed interviews, one moderator and two evaluators were assigned to each interview. The moderator was always one of the interviewers. The first evaluator was the other interviewer, while the second evaluator was one of the authors who did not participate in the interview. The role of each evaluator was to perform the open coding task. In contrast, the moderator's role was to identify and resolve inconsistently-labelled fragments of text between the evaluators (\eg different tags attached to the same fragment of text). We decided to involve the interviewers in this task in two roles (evaluator and moderator) because they were more informed of the content and context of the interview, having been exposed to the informal and meta aspects of the communication with the interviewees. The second evaluator who was not involved in the interview ensured the presence of a different point of view. 

Twenty interviews were equally divided among the authors who did not participate in the interview process. Each interviewer was the evaluator of 10 interviews and the moderator for the remaining 10. Overall, 297 pieces of text were tagged by the evaluators. Among them, there were only 6 cases of conflict, where the evaluators attached different tags to the same fragment of text. Moreover, there were 196 cases when one evaluator put a tag on a fragment of text, while the other did not. Among these cases, 146 were kept by the moderators, while the rest were discarded. As a result of this process, 245 final tags were extracted. The number of tags per interview ranged between 5 and 22, with an average of 12 tags. 

Once the open coding of all interviews was completed, a final meeting with all the authors took place. At this meeting, authors went through the final list of tags, focusing in particular on the tags that were deemed not related to issues and problems in DL systems, but rather had a more general nature. After this discussion, 19 tags were removed, leaving the final 226 tags available for the taxonomy. 

\subsection{Taxonomy Construction and Validation}

To build the taxonomy we used a bottom-up approach \cite{Vijayaraghavan:2003}, where we first grouped tags that correspond to similar notions into categories. Then, we created parent categories, ensuring that categories and their subcategories follow an ``is a'' relationship. Each version of the taxonomy was discussed and updated by all authors in the physical meetings associated with the tagging rounds. At the end of the construction process, in a physical meeting the authors went together through all the categories, subcategories and leaves of the final taxonomy for the final (minor) adjustments. 

To ensure that the final taxonomy is comprehensive and representative of real DL faults, we validated it through a  survey involving a new set of practitioners/researchers, different from those who participated in the interviews. 

To recruit candidates for the survey, we adopted the same strategy and selection criteria as the one we used for the interview process (\secref{sec:int_recr}). The first group of candidates we contacted was derived from authors' personal contacts. We contacted 23 individuals remaining from our initial list and 13 of them actually filled the survey. The second and the third group of candidates came from SO and Upwork, respectively. From the SO platform we selected the top 20 answerers from the `Last 30 Days' and `All time' categories for each of the three considered frameworks. By the time we were performing the survey, these two categories had partly changed in terms of the featured users, so there were new users also in the top 10 lists. From the set of 120 users we discarded those who had been already contacted for the interviews. Among the remaining candidates, we were able to access contact details of only 20 users. We contacted all of them and 4 have completed the survey. For the Upwork group, we created a new job posting with a fixed payment of 10 USD per job completion and sent an offer to 26 users. Four of them completed the survey. Overall, 21 participants took part in our survey (10 researchers and 11 practitioners), with a minimum overall coding experience of 1 year and a maximum of 20 years (median=4). Concerning the relevant DL experience, the minimum was 1 year and the maximum 7 years (median=3).

To create our survey form we used Qualtrics \cite{qualtrics}, a web-based tool to conduct survey research, evaluations and other data collection activities. We started the survey with the same background questions as in our interviews. Then, we proceeded with the questions related to our final taxonomy. Putting the whole taxonomy structure in a single figure of the survey would make it overly complicated to read and understand. Therefore, we partitioned it by inner categories, choosing either the topmost inner category, when it was not too large, or its descendants, when it was a large one. 

For each inner category that partitioned the taxonomy, we created a textual description including examples of its leaf tags. In the survey form, we presented the name of the category, its textual description, and then three questions associated with it. The first question was a ``yes'' or ``no'' question on whether the participant had ever encountered this problem. In case of positive answer, we had two more Likert-scale questions on the severity of the issue and the amount of  effort required to identify and fix it. In this way we evaluated not only the mere occurrence of a taxonomy fault, but also its severity as perceived by developers. 

In the final part of our survey, we asked the participants to list problems related to DL that they have encountered, but which had not been mentioned in the survey. By doing this we could check whether our taxonomy covered all the faults in developer's experience, and if it does not, we could find out what is missing.

%% file: results.tex

\section{Results} \label{sec:results}

The material used to conduct our study and the (anonymized) collected data are publicly available for replication purposes \cite{replication}. 

\subsection{The Final Taxonomy} \label{sec:taxonomy}

\begin{figure*}[htb]
\begin{center}
\includegraphics[angle=90, width=0.965\textwidth]{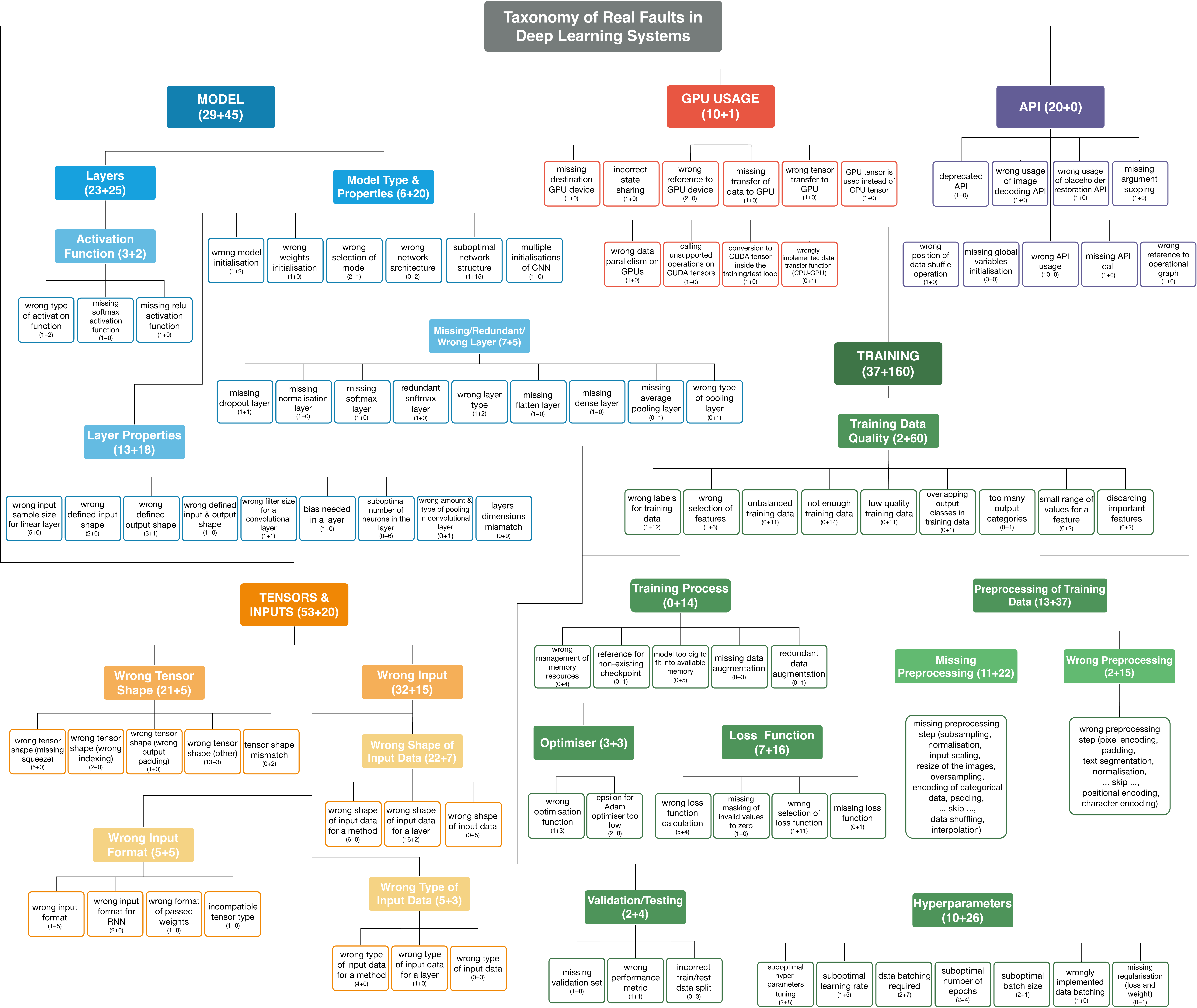}
\caption{Final Taxonomy}
\label{fig:taxonomy}
\end{center}
\end{figure*} 

The taxonomy is organised into 5 top level categories, 3 of which are further divided into inner subcategories. The full taxonomy is shown in \figref{fig:taxonomy}. The two numbers separated by a plus sign after each category name represent the number of posts assigned to such a category during manual labelling and the number of occurrences of such a tag in the interviews after open coding, respectively.

\textbf{Model.} This category of the taxonomy covers faults related to the structure and properties of a DL model.

\textit{Model Type \& Properties.}
This category considers faults affecting the model as a whole, rather than its individual aspects/components. One such fault is a wrong selection of the model type, for example, when a recurrent network was used instead of a convolutional network for a task that required the latter. In addition, there are several cases of incorrect initialisation of a model, which result in the instability of the gradients. 
Another common pitfall from this category is using too few or too many layers, causing suboptimal network structure, which in turn leads to poor performance of the model. An example was provided by one of our interviewees: "\textit{when we started, we were thinking that we needed at least four layers in the encoder and the decoder and then we ended up having half of them, like actually very shallow model and it was even better than the bigger deeper model}". 

\textit{Layers.} Faults in this category affect a particular layer of a neural network. This is a large taxonomy category that was further divided into the three inner subcategories described below:

\begin{itemize}[leftmargin=*]
  \item \textit{Missing/Redundant/Wrong Layer.} These faults represent cases where adding, removing or changing the type of a specific layer was needed to remedy the low accuracy of a network. This is different from the \textit{suboptimal network structure} of \textit{Model Type \& Properties} category, as here the solution is local to a specific layer, rather than affecting the whole Smodel. An interviewee described such a fault, which was related "\textit{not to the wrong architecture as whole, but more usually to the wrong type of layer, because usually in our field people have applied type of layers which were not suited for the type of input which they are processing}". 

 \item \textit{Layer Properties.} This category represents faults due to some layer's incorrect inner properties, such as its input/output shape, input sample size, number of neurons in it. As per interviewee's description, "\textit{we set too large number of neurons and we had like very slow training and validation}".
 
 \item \textit{Activation Function.} Another important aspect of a neural network is the activation function of neurons. If not selected properly, it can dramatically ruin the model's performance. One interviewee noted that "\textit{when I changed sigmoid activations into linear activations in the speech recognition, it gave me a gain}". 
 
\end{itemize}

\textbf{Tensors \& Inputs.} This category deals with problems related to the wrong shape, type or format of the data. We encountered two different classes of faults in this category:

\textit{Wrong Tensor Shape.} A faulty behaviour manifests during some operation on tensors with incompatible shapes or on a single tensor with incorrectly defined shape.\HIDE{ As shown in  \figref{fig:taxonomy}, there are several causes for a wrong tensor shape, \eg missing output padding, missing indexing, or a case when an interviewee "\textit{was using a transposed version of the tensor instead of the normal one}".} As shown in  \figref{fig:taxonomy}, there is a number of possible causes for a wrong tensor shape, \eg missing output padding, missing indexing, or, as it was provided in one of the interviews, a case when a developer "\textit{was using a transposed version of the tensor instead of the normal one}".

\textit{Wrong Input.} A faulty behaviour is due to data with incompatible format, type or shape being used as an input to a layer or a method. A wrong input to a method is a problem  frequently observed in traditional software, as well as in DL programming. However, in DL, these faults happen to be of a specific nature, ranging from the input having unexpected datatype (\eg \textit{string} instead of \textit{float}) or shape (a tensor of size \textit{5x5} instead of \textit{5x10)} to cases when the input has a completely wrong format (\eg a wrong data structure). One interesting example of a wrong input format was provided by our interviewee: "\textit{my data was being loaded in with channel access first instead of last. So that actually was a silent bug and it was running and I actually don't understand how it even ran but it did}".

\textbf{Training.} This is the largest category in the taxonomy and it includes a wide range of issues related to all facets of the training process, such as the quality and preprocessing of training data, tuning of hyperparameters, the choice of appropriate loss/optimisation function. It also accounts for the faults occurring when testing/validating a previously trained model.

\textit{Hyperparameters.} Developers face a large number of problems when tuning the hyperparameters of a DL model. The most reported incorrect hyperparameters are learning rate, databatch size, and number of epochs. While suboptimal values for these parameters do not necessarily lead to a crash or an error, they can affect the training time and the overall performance achieved by the model. An example from the interviews is "\textit{when changing learning rate from 1 or 2 orders of magnitude, we have found that it impacts the performance of about up to 10\% to 15\% in terms of accuracy}".

\textit{Loss Function.} This category contains faults associated with the loss function, specifically its selection and calculation. Wrong selection of the loss function or usage of a predefined loss function may not adequately represent the optimisation goals that a model is expected to achieve. In its turn, a wrong calculation of a loss function occurs when a custom loss function is implemented and some error in the implementation leads to the suboptimal or faulty behaviour. As one interviewee noted, they needed "\textit{to get a more balanced loss function than just something that can predict one class very well and then screws up the other ones}".

\textit{Validation/Testing.} It includes problems related to testing and validating a trained model, such as the bad choice of performance metrics or faulty split of data into training and testing datasets.

\textit{Preprocessing of Training Data.} Preprocessing of a training dataset is a labour-intensive process that significantly affects the performance of a DL system. This is reflected in the large number of elements in this category and in the high variety and number of its leaves. At the high-level we have separated the faults in this category into two groups: missing preprocessing and wrong preprocessing. The former refers to cases when a preprocessing step that would lead to a better performance has not been applied at all. In the latter case, the preprocessing step has actually been applied, but either it was of an unsuitable type or was applied in an incorrect way. Examples of the most frequent issues are missing normalisation step, missing input scaling or subsampling, and wrong pixel encoding. It is important to remark that preprocessing steps for training data are heavily dependent on an area of application. This explains the large variety of leaf tags in this category. We had to omit some of them from the taxonomy figure, due to the lack of space.

\textit{Optimiser.} This category is related to the selection of an unsuitable optimisation function for model training. Wrong selection of the optimiser (\eg Adam optimiser instead of stochastic gradient descent) or suboptimal tuning of its parameters (\textit{too low epsilon for Adam optimiser}) can ruin the performance of a model.

\textit{Training Data Quality.} In this group fall all the aspects relevant to the quality of training data. In general, issues occur due to the complexity of the data and the need for manual effort to ensure a high quality of training data (\eg to label and clean the data, to remove the outliers). More specific cases of data collection challenges include privacy issues in the medical field and constantly changing user interfaces of web pages, \HIDE{ \NARGIZ{I do not remember that interfaces were leading to not enough training data. mb just a challenge? or because of that there might be some unrelated information crawled, but I can be mistaken}}from which the data is gathered automatically. All of this leads to the most frequent issue in this category, which is \textit{not enough training data}. A variant of this problem is \textit{unbalanced training data}, where one or more classes in a dataset are underrepresented. Moreover, to get a good classification model, it is important to ensure the provision of correct labels for training data.  However, in the interviewees' experience getting \textit{wrong labels for training data} is a common and an annoying issue. The set of other issues related to the quality of training data, such as the lack of a standard format, missing pieces of data or the presence of unrelated data (\eg images from other domains) are gathered together under a rather general tag \textit{low quality of training data}, because specific issues depend on the area of application.

\textit{Training Process.} This category represents the faults developers face during the process of model training, such as \textit{wrong management of memory resources} or \textit{missing data augmentation}. It also contains leaves representing the exploitation of models that are too big to be fitted into available memory or reference to non-existing checkpoints during model restoration. Regarding the data augmentation, one of the interviewees noted that it helped "\textit{to make the data more realistic to work better in low light environments"}, while the other said that sometimes "\textit{you add more pictures to data set}" and as a result you can face "\textit{the overfitting of the network problem, so sometimes data augmentation can help, sometimes it can damage}". 

\textbf{GPU Usage.} This top-level category gathers all kinds of faults related to the usage of GPU devices while working with DL. There is no further division in this case as all the examples we found represent very specific issues. Some highlights from this category are: \textit{wrong reference to GPU device}, \textit{failed parallelism}, \textit{incorrect state sharing between subprocesses}, \textit{faulty transfer of data to a GPU device}.

\textbf{API.} This part of the taxonomy represents a broad category of problems arising from framework's API usage. The most frequent is \textit{wrong API usage}, which means that a developer is using an API in a way that does not conform to the logic set out by developers of the framework. Another illustrating example could be a missing or wrongly positioned API call.

\vspace{-0.15cm}
\subsection{Contributions to the Taxonomy}

The final taxonomy was built using tags extracted from two different sources of information: SO \& GitHub artifacts and researcher/practitioner interviews. The top 5 tags obtained from SO \& GitHub with their respective number of occurrences (shown as \textit{NN + MM}, where \textit{NN} refers to SO \& GitHub; \textit{MM} to interviews) are \textit{wrong tensor shape} (21+5), \textit{wrong shape of input data for a layer} (16+2), \textit{missing preprocessing} (11+22), \textit{wrong API usage} (10+0) and \textit{wrong shape of input data for a method} (6+0). 

For the interviews, the top 5 tags are \textit{missing preprocessing} (11+22), \textit{suboptimal network structure} (1+15), \textit{wrong preprocessing} (2+15), \textit{not enough training data} (0+14) and \textit{wrong labels for training data} (1+12). These lists have an intersection of only one tag (\textit{missing preprocessing}). The top 5 SO \& GitHub list contains two tags that did not occur in the other source (\textit{wrong API usage}, \textit{wrong shape of input data for a method}). The top 5 interview list contains one such tag (\textit{not enough training data}). Moreover, the number of occurrences is unbalanced between the two sources: for the top 5 SO \& GitHub tags, there are 64+29 occurrences, while for the top 5 interview tags the number becomes 15+78. This shows that the two selected sources are quite complementary. 

Indeed, the complementarity between these sources of information is reflected in the overall taxonomy structure. If we consider the five top level categories in the taxonomy (\ie the five direct children of the root node in the taxonomy), we can find one category to which  interview tags have not contributed at all, namely, the \textit{API} category. This might be due to the fact that API-related problems are specific and therefore, they did not come up during interviews, where interviewees tended to talk about more general problems. Similarly, in the \textit{GPU Usage} category there is only one interview tag. \textit{Tensors \& Inputs} is another category dominated by SO \& GitHub tags, the number of which is twice the number of interview tags. In contrast, the main contributors to the \textit{Model} category are  interviews. The largest difference is for the \textit{Training} category, where  interviews contributed 4 times more tags, which led to addition of two more subcategories. The presence of training related faults only in the interviews is expected, as these types of problems can not usually be solved by asking a question on SO or opening an issue on GitHub. Out of 18  pre-leaf categories, one consists of  tags provided only by SO \& GitHub (\textit{API}) and another one only by interviews (\textit{Training Process}). Another pre-leaf category (\textit{Training Data Quality}) was abstracted from the few existing leaves only after collecting more data from the interviews. The remaining 16 consist of different proportions of the two sources, with 6 having higher number of SO \& GitHub tags, 8  higher number of interview tags and 2  the same amount of tags from the two sources.

Overall, the distribution of the tags shows that SO \& GitHub artefacts and researcher/practitioner interviews are two very different and complementary sources of information. Ignoring one of them would provide an incomplete taxonomy, which would not be representative of the full spectrum of real DL faults.

\vspace{-0.1cm}
\subsection{Validation Results}

The results of the validation survey are summarised in \tabref{tab:validation}. For each category, we report the percentage of ``yes'' and ``no'' answers to the question asking whether participants ever encountered the related issues. We also show the perceived severity of each fault category and the perceived effort required to identify and fix faults in such a category. 
There is no category of faults that the survey participants have never encountered in their experience, which confirms that all the categories in the taxonomy are relevant. The most approved category is \textit{Training Data}, with 95\% of ``yes'' answers. According to the respondents, this category has ``Critical'' severity and  requires ``High'' effort for 61\% and 78\% of participants,  respectively. The least approved category is \textit{Missing/Redundant/Wrong Layer}, which has been experienced by  24\% of the survey participants (a non negligible fraction of all the participants). Across all the categories, the average rate of ``yes'' answers is 66\%, showing that the final taxonomy contains categories that match the experience of a large majority of the participants (only two categories are below 50\%). Participants confirmed, on average, 9.7 categories (out of 15) across all the surveys.

\renewcommand{\tabcolsep}{2pt}
\begin{table}[htb]
\scriptsize
\caption{Validation Survey Results}
\vspace{-0.3cm}
\begin{tabular}{l|cc|ccc|ccc}
\hline
\multirow{2}{*}{\textbf{Category}} & \multicolumn{2}{c|}{\textbf{Response}} & \multicolumn{3}{c|}{\textbf{Severity}} & \multicolumn{3}{c}{\textbf{Effort Required}}\\
& \multicolumn{1}{c}{Yes}  & \multicolumn{1}{c|}{No}  & \multicolumn{1}{c}{Minor}  & \multicolumn{1}{c}{Major}  & \multicolumn{1}{c|}{Critical}  & \multicolumn{1}{c}{Low} &  \multicolumn{1}{c}{Medium} & \multicolumn{1}{c}{High} \\
\hline
Hyperparameters & 86\% & 14\% & 44\% & 44\% & 11\% & 22\% & 33\% & 44\%\\
Loss Function & 65\% & 35\% & 15\% & 54\% & 31\% & 23\% & 46\% & 31\%\\
Validation \& Testing & 60\% & 40\% & 33\% & 42\% & 25\% & 50\% & 17\% & 33\%\\
Preprocessing of Training Data & 86\% & 14\%  & 56\% & 17\% & 28\% & 28\% & 39\% & 33\%\\
Optimiser & 57\% & 43\% & 75\% & 25\% & 0\% & 58\% & 33\% & 8\%\\
Training Data & 95\%  & 5\% & 6\% & 33\% & 61\% & 6\% & 17\% & 78\%\\
Training Process & 68\% & 32\% & 31\% & 23\% & 46\% & 31\% & 15\% & 54\%\\
Model Type \& Properties & 81\%  & 19\% & 44\% & 44\% & 13\% & 38\% & 44\% & 19\%\\
Missing/Redundant/Wrong Layer & 24\% & 76\% & 60\% & 40\% & 0\% & 60\% & 40\% & 0\%\\
Layer Properties & 76\%  & 24\% & 44\% & 44\% & 13\% & 63\% & 31\% & 6\%\\
Activation Function & 43\%  & 57\% & 33\% & 67\% & 0\% & 67\% & 22\% & 11\%\\
Wrong Input & 62\% & 38\% & 62\% & 31\% & 8\% & 69\% & 31\% & 0\%\\
Wrong Tensor Shape & 67\%  & 32\% & 57\% & 21\% & 21\% & 71\% & 29\% & 0\%\\
GPU Usage & 52\%  & 48\% & 55\% & 18\% & 27\% & 55\% & 18\% & 27\%\\
API & 67\% & 33\% & 43\% & 29\% & 29\% & 36\% & 43\% & 21\%\\
\hline
\end{tabular}
\label{tab:validation}
\vspace{-0.3cm}
\end{table}

Some participants provided examples of faults they thought were not part of the presented taxonomy. Three of them were generic coding problems, while one participant described the effect of the fault, rather than its cause. The remaining three could actually be placed in our taxonomy under \textit{"missing API call"}, \textit{"wrong management of memory resources"} and \textit{"wrong selection of features"}.  We think the participants were not able to locate the appropriate category in the taxonomy because the descriptions in the survey did not include enough exemplar cases, matching their specific experience.

%% file: discussion.tex

\section{Discussion} \label{sec:discussion}

\textbf{Final Taxonomy vs. Related Work.}  
To elaborate on the comparison with existing literature, we analysed the differences between our taxonomy and the taxonomy from the only work where authors compiled their own classification of faults, rather than reusing an existing one, which is by Zhang \etal \cite{Zhang:2018}. To ease the comprehension, we list the categories with the exact naming and excerpts of descriptions from the corresponding publication \cite{Zhang:2018}:

\textit{1. Incorrect Model Parameter or Structure (IPS)} - ``bugs related to modelling mistakes arose from either an inappropriate model parameter like learning rate or an incorrect model structure like missing nodes or layers''.
In our taxonomy we distinguish the selection of an appropriate model structure from the tuning of the hyperparameters. So, in our taxonomy, class IPS corresponds to two leaves: \textit{suboptimal network structure} and \textit{suboptimal hyperparameters tuning}, each belonging to a different top-level category -- \textit{Model} and \textit{Training}, respectively.

\textit{2. Unaligned Tensor (UT)} - ``a bug spotted in computation graph construction phase when the shape of the input tensor does not match what it is expected''. Class UT can be mapped to the \textit{Wrong Tensor Shape} and partly to the \textit{Wrong Shape of Input Data} (as far as it concerns tensors) categories of our taxonomy.

\textit{3. Confusion with TensorFlow Computation Model (CCM)} - ``bugs arise when TF users are not familiar with the underlaying computation model assumed by TensorFlow''. 
CCM deals with the data flow semantics of tensors, which might be unintuitive to novices. This is a difficulty that developers face when starting to work with tensors in general. We did not gather evidence for this fault because we excluded examples, toy programs, and tutorials from our analysis. 

Zhang \etal did not observe this fault in GitHub (only in SO). As they remark: ``they can be common mistakes made by TF users and discussed at Stack Overflow seeking advice''.

\textit{4. Tensor Flow API Change (APIC)} - ``anomalies can be exhibited by a TF program upon a new release of TensorFlow libraries''. As these bugs are related to the evolution of the framework, they are similar to those that affect any code using third party libraries. Hence, we regarded them as generic programming bugs, not DL-specific faults. 

\textit{5. TensorFlow API Misuse (APIM)} - ``bugs were introduced by TF users who did not fully understand the assumptions made by the APIs''. This  class can be directly linked to the \textit{wrong API usage} leaf in the \textit{API} category, with the only difference that in our case the leaf includes APIs from three, not just one, DL frameworks.

\textit{6. Structure Inefficiency (SI)} - ``a major difference between SI and IPS is that the SI leads to performance inefficiency while the IPS leads to functional incorrectness''. This class of bugs is similar to class \textit{IPS}, differing only in the observable effects (functional vs efficiency problems), which are not taken into account in our taxonomy (we looked at the root cause of a fault, not at its effects). So, the mapping is the same as for IPS.

\textit{7. Others (O)} - ``other bugs that cannot be classified are included in this type. These bugs are usually programming mistakes unrelated to TensorFlow, such as Python programming errors or data preprocessing''. Generic programming errors are excluded from our taxonomy, which is focused on DL-specific faults. Data preprocessing errors instead correspond to the category \textit{Preprocessing of Training Data}.

In summary, Zhang \etal's classes IPS, SI and APIM map to 3 leaf nodes of our taxonomy; classes UT and O map to 3 inner nodes (although for 2 out of 3 the mapping is partial). In total (see Table~\ref{tab:labeling}), our taxonomy has 24 inner nodes and 92 leaf nodes. So, our taxonomy contains 21 inner nodes (out of 24) that represent new fault categories with respect to Zhang \etal's (19 out of 24 if we do not count descendants of mapped nodes). For what concerns the leaves, the computation is more difficult when the mapping is partial, because it is not always easy to decide which subset of leaves is covered by classes UT and O. If we conservatively overestimate that all leaves that descend from a partially mapped node are transitively covered,  Zhang \etal's classes would cover 13 leaf nodes, out of 92, in our taxonomy. This means that 79 leaf categories have been discovered uniquely and only in our study. On the other hand, the two unmapped classes by Zhang \etal (CCM and APIC) correspond to generic programming bugs or bugs faced by novices who seek advice about tensor computation in SO. We deliberately excluded such classes from our analysis.

Overall, a large proportion of inner fault categories and of leaf fault categories in our taxonomy are new and unique to our study, which hence represents a substantial advancement of the knowledge of real DL faults over the previous work by Zhang \etal

\textbf{Final Taxonomy vs. Mutation Operators.} Mutants are artificial faults that are seeded into a program to test under the assumption that fault revelation will translate from mutants to real faults. The works by Ma \etal \cite{Ma:2018} and Shen \etal \cite{Shen:2018} have made initial attempts to define mutation operators for DL systems. The combined list of mutation operators from these works can be classified into two categories: (1) \textit{Pre-Training Mutations}, applied to the training data or to the model structure before training is performed; (2) \textit{Post-Training Mutations} that change the weights, biases or structure of a model that has already been trained. For each pre-training mutant, after mutation the model must be retrained, while for post-training mutants no retraining is needed. 

Whether mutants are a valid substitute for real faults has been an ongoing debate for traditional software \cite{Andrews:2005,Daran:1996,Just:2014}. To obtain an insight on the correspondence between the proposed DL mutants and real faults in DL systems, we matched the mutation operators from the literature to the faults in our taxonomy. \tabref{tab:mutation} lists each pre-training mutation operator and provides the corresponding taxonomy category when such a match exists. This is the case for all pre-training mutation operators except ``Data Shuffle''.

\vspace{-0.1cm}
\begin{table}[ht]
\scriptsize
\caption{Taxonomy Tags for Mutation Operators from \cite{Ma:2018}}
\vspace{-0.25cm}
\begin{tabular}{l|l} 
\hline
\bf Mutation Operator & \bf Taxonomy Category \\
\hline
Data Repetition & Unbalanced training data\\
Label Error & Wrong labels for training data \\
Data Missing & Not enough training data\\
Data Shuffle & - \\
Noise Perturbation & Low quality of training data\\
Layer Removal & Missing/redundant/wrong layer\\
Layer Addition & Missing/redundant/wrong layer\\
Activation Function Removal & Missing activation function\\
\hline
\end{tabular}
\vspace{-0.15cm}
\label{tab:mutation}
\end{table}

For what concerns the post-training mutants, there is no single fault in the taxonomy related to the change of model parameters after the model has already been trained. Indeed, these mutation operators are very artificial and we assume they have been proposed as they do not require retraining, \ie are cheaper to generate. However, their effectiveness is still to be demonstrated.

Overall, we can notice that the existing mutation operators do not capture the whole variety of real faults present in our taxonomy, as out of 92 unique real faults (leaf nodes) from the taxonomy, only 6 have a corresponding mutation operator. While some taxonomy categories may not be suitable to be turned into mutation operators, we think that there is ample room for the design of novel mutation operators for DL systems based on the outcome of our study.

\textbf{Stack Overflow \emph{vs.} Real World.}  The study by Meldrum \etal \cite{Meldrum:2017}, which analyses 226 papers that use SO, demonstrates the growing impact of SO on software engineering research. However, the authors note that this raises quality-related concerns, as the utility and reliability of SO is not validated in any way. Indeed, we collected feedback on this issue when interviewing the top SO answerers.
All SO interviewees agreed that the questions asked on SO are not entirely representative of the problems developers encounter when working on a DL project. One interviewee noted that these questions are ``\emph{somehow different from the questions my colleagues will ask me}'', while the other called them ``\emph{two worlds, completely different worlds}". Interviewees further elaborated on why they think this difference exists. One reasoning was that ``\emph{most engineers in the industry have more experience in implementing, in tracing the code}'', so their problems are not like ``\emph{how should I stack these layers to make a valid model, but most questions on SO are like model building or why does it diverge kind of questions}''. Another argument was that the questions on SO are mostly  ``\emph{sort of beginner questions of people that don't really understand the documentation}'' and that they are asked by people who ``\emph{are extremely new to linear algebra and to neural networks}''. 

We addressed this issue by excluding examples, toy programs and tutorials from the set of analysed artefacts and by complementing our taxonomy with developer interviews.

\textbf{Common Problems.} Our interviews with developers were conducted to get information on DL faults. However, due to the semi-structured nature of these interviews, we ended up collecting information on more topics than that. The \textit{version incompatibility} between different libraries and frameworks was one of interviewees' main concerns. They also expressed their dissatisfaction with the \textit{quality of documentation} available, with one developer noting that this problem is even bigger for non computer vision problems. Another family of problems mentioned very often was the \textit{limited support of DL frameworks} for a number of tasks, such as implementation of custom loss functions and of custom layers, serialisation of models, and optimisation of model structure for complex networks. The \textit{lack of tools} to support activities such as performance evaluation, combining outputs of multiple models, converting models from one framework to another was yet another challenging factor according to our interviewees.

%% file: threats.tex

\section{Threats to Validity} \label{sec:threats}

\textbf{Internal}. A threat to the internal validity of the study could be the biased tagging of the artefacts from SO \& GitHub, and of the  interviews. To mitigate this threat, each artifact and interview was labelled by at least two evaluators. Also, it is possible that questions asked during the developer interviews might have been affected by the initial taxonomy based on SO \& GitHub tags or that they have directed the interviewees towards specific types of faults. To prevent this from happening, we kept the questions as generic and disjoint from the initial taxonomy as possible. Another threat might be related to the procedure of building the taxonomy structure from a set of tags. As there is no unique and correct way to perform this task, the final structure might have been affected by the authors' point of view. For this reason it was validated via survey.

\textbf{External}. The main threat to the external validity is generalisation beyond the three considered frameworks, the dataset of artefacts used and the interviews conducted. We selected the frameworks based on their popularity. Our selection was further confirmed by the list of frameworks that developers from both the interviews and survey had used in their experience. To make the final taxonomy as comprehensive as possible, we labeled a large number of artefacts from SO \& GitHub until we reached saturation of the inner categories. To get diverse perspectives from the interviews, we recruited developers with different levels of expertise and background, across a wide range of domains. 


%% file: conclusion.tex

\section{Conclusion} \label{sec:conclusion}

We have constructed a taxonomy of real DL faults, based on manual analysis of 1,059 GitHub \& SO artefacts and interviews with 20 developers. The taxonomy is composed of 5 main categories containing 375 instances of 92 unique types of faults. To validate the taxonomy, we conducted a survey with a different set of 21 developers who confirmed the relevance and completeness of the identified categories. In our future work we plan to use the presented taxonomy as a guidance to improve DL systems testing and as a source for the definition of novel mutation operators. 


\HIDE{The percentage of responses confirming that they have encountered a fault from a specific subcategory is 65.87\%}